# Electronics of Time-of-flight Measurement for Back-n at CSNS

T. Yu, P. Cao, X.Y. Ji, L.K. Xie, X.R. Huang, Q. An, H.Y. Bai, J. Bao, Y.H. Chen, P.J. Cheng, Z.Q. Cui, R.R. Fan, C.Q. Feng, M.H. Gu, Z.J. Han, G.Z. He, Y.C. He, Y.F. He, H.X. Huang, W.L. Huang, X.L. Ji, H.Y. Jiang, W. Jiang, H.Y. Jing, L. Kang, B. Li, L. Li, Q. Li, X. Li, Y. Li, R. Liu, S.B. Liu, X.Y. Liu, G.Y. Luan, Y.L. Ma, C.J. Ning, J. Ren, X.C. Ruan, Z.H. Song, H. Sun, X.Y. Sun, Z.J. Sun, Z.X. Tan, J.Y. Tang, H.Q. Tang, P.C. Wang, Q. Wang, T.F. Wang, Y.F. Wang, Z.H. Wang, Z. Wang, J. Wen, Z.W. Wen, Q.B. Wu, X.G. Wu, X. Wu, Y.W. Yang, H. Yi, L. Yu, Y.J. Yu, G.H. Zhang, L.Y. Zhang, J. Zhang, Q.M. Zhang, Q.W. Zhang, X.P. Zhang, Y.T. Zhao, Q.P. Zhong, L. Zhou, Z.Y. Zhou and K.J. Zhu

*Abstract*—Back-n is a white neutron experimental facility at China Spallation Neutron Source (CSNS). The time structure of the primary proton beam make it fully applicable to use TOF (time-of-flight) method for neutron energy measuring. We implement the electronics of TOF measurement on the general-purpose readout electronics designed for all of the seven detectors in Back-n. The electronics is based on PXIe (Peripheral Component Interconnect Express eXtensions for Instrumentation) platform, which is composed of FDM (Field Digitizer Modules), TCM (Trigger and Clock Module), and SCM (Signal Conditioning Module). T0 signal synchronous to the CSNS accelerator represents the neutron emission from the target. It is the start of time stamp. The trigger and clock module (TCM) receives, synchronizes and distributes the T0 signal to each FDM based on the PXIe backplane bus. Meantime, detector signals after being conditioned are fed into FDMs for waveform digitizing. First sample point of the signal is the stop of time stamp. According to the start, stop time stamp and the time of signal over threshold, the total TOF can be obtained. FPGA-based (Field Programmable Gate Array) TDC is implemented on TCM to accurately acquire the time interval between the asynchronous T0 signal and the global synchronous clock phase. There is also an FPGA-based TDC on FDM to accurately acquire the time interval between T0 arriving at FDM and the first sample point of the detector signal, the over threshold time of signal is obtained offline. This method for TOF measurement is efficient and not needed for additional modules. Test result shows the accuracy of TOF is sub-nanosecond and can meet the requirement for Back-n at CSNS.

## I. INTRODUCTION

THE China Spallation Neutron Source (CSNS) is a large scientific facility designed for multidisciplinary research on material characterization using neutron scattering techniques [1].The first phase (CSNS-I) is 100kW in beam power and the upgrading phase (CSNS-II) will be increased to 500kW. With the beam energy up to 1.6GeV and 25 Hz in pulse repetition rate, the CSNS accelerator is a powerful facility to serve researches in neutron beam applications [2].

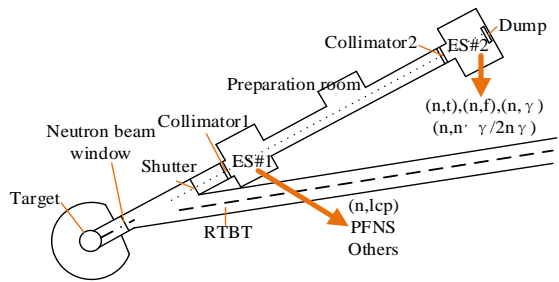

Fig. 1. The layout of WNS beam line and end stations

The proton beam line RTBT in CSNS have a 15° bending at a distance of about 20 m from the spallation target. The back-

Manuscript submitted June 24, 2018.This work is supported by National Research and Development plan (No. 2016YFA0401602) and NSAF (No.U153011).

T. Yu, X.R. Huang, Q. An, C.Q. Feng, S.B. Liu and L. Yu are with the Department of Modern Physics, University of Science and Technology of China, Hefei, 230026, China.

P. Cao, X.Y. Ji and L.K. Xie are with the Department of Engineering and Applied Physics, University of Science and Technology of China, Hefei 230026, China. (Corresponding author: Ping Cao, e-mail: cping@ustc.edu.cn).

T. Yu, P. Cao, X.Y. Ji, L.K. Xie, X.R. Huang, Q. An, R.R. Fan, C.Q. Feng, S.B. Liu, Z.J. Sun, L. Yu and L.Y. Zhang are with the State Key Laboratory of Particle Detection and Electronics, Beijing 100049, Hefei 230026, China

T.F. Wang is with Beihang University, Beijing, 100083, China.

J. Bao, G.Z. He, H.X. Huang, G.Y. Luan, J. Ren, X.C. Ruan, H.Q. Tang, Q. Wang, Z.H. Wang, X.G. Wu, Q.W. Zhang, Q.P. Zhong and Z.Y. Zhou are with the China Institute of Atomic Energy, Beijing, 102413, China.

Y.H. Chen, R.R. Fan, Y.C. He, W.L. Huang, W. Jiang, H.T. Jing, L. Kang, B. Li, L. Li, Q. Li, X. Li, Y.L. Ma, C.J. Ning, H. Sun, X.Y. Sun, Z.J. Sun, Z.X. Tan, J.Y. Tang, P.C. Wang, Y.F. Wang, Z. Wang, Q.B. Wu, X. Wu, H. Yi, Y.J. Yu, L.Y. Zhang, J. Zhang and L. Zhou are with the Institute of High Energy Physics, CAS, Beijing 100049, China and Dongguan Neutron Science Center, Dongguan 523803, China.

M.H. Gu, X.L. Ji, Y. Li and K.J. Zhu are with the Institute of High Energy Physics, CAS, Beijing 100049, China.

Z.J. Han, R. Liu, X.Y. Liu, J. Wen, Z.W. Wen and Y.W. Yang are with the Institute of Nuclear Physics and Chemistry, CAEP, Mianyang, China.

Z.H. Song and X.P. Zhang are with the Northwest Institute of Nuclear Technology, Xi'an, China.

H.Y. Bai, Z.Q. Cui, H.Y. Jiang and G.H. Zhang are with the Peking University, Beijing, 100871, China.

P.J. Cheng and Y.F. He are with the University of South China, Hengyang, 421001, China.

Q.M. Zhang and Y.T. Zhao are with the Xi'an Jiaotong University, Xi'an, 710049, China.

streaming neutrons (Back-n) from the spallation target are departed from the charged particles by the magnet on the bending, studies show that it's suitable to exploit the back-streaming as a white neutron source for nuclear data measurements [3][4]. Fig.1 is the layout of Back-n beam line and end stations.

There will be seven spectrometers in Back-n: $C_6D_6$ detectors with 4 detector units for cross section (n, γ) early measurements, a $4\pi$-$BaF_2$ array named GTAF-II (Gamma Total Absorption Facility II) for neutron capture cross section (n, γ) measurements, FIXM (Fast Ionization Chamber Spectrometer for Fission Cross section Measurement) with eight-layer fast ionization chamber for fission cross section (n, f) measurements, LPDA (Light-charged Particle Detector Array) with a light-charged particle emission (n, lcp) detector array, FINDA (Fission Neutron Spectrum Detector Array) with a detector array for prompt fission neutron spectrum (PFNS) measurements, and GAEA (Gamma Spectrometer with Germanium Array) with a $4\pi$-germanium detector array for gamma spectrum (n, n'γ/2nγ) measurements. For all these detectors, the measurement of neutron energy is important.

In the case of neutron time-of-flight (n_TOF), triggerless technical route is adopted, flash-ADCs with 8-bit quantization accuracy and 500 MSPS typical sampling rate are used to digitize the signals, the rate of neutron pulse is 0.8Hz or lower, the CPCI (Compact Peripheral Component Interconnect) chassis is enough to upload continue data without trigger for 8 to 10 channels [5-9]. In Back-n, GTAF-II has the strictest requirements for readout electronics in all the seven detectors. The signals of GTAT-II are range from 4mV to 2000mV, to obtain signals more accurately, ADCs with 1 GSPS sampling rate and 12-bit quantization accuracy are used. The high sampling rate and high accuracy increase the readout data, the rate of neutron pulse in Back-n is 25Hz which increase the readout data a step further. Obviously the CPCI platform with triggerless technical route is not enough for data readout at Back-n, the general-purpose readout electronics based on PXIe (Peripheral Component Interconnect Express eXtensions for Instrumentation) platform which owns high speed serial bus is adopted [10]. The trigger system is used to lower the readout data, the time stamp is used to reestablish signals in one time axis offline, see chapter II for details.

Time of flight (TOF) method is an important way to measure neutron energy. In the electronics of TOF measurement, T0 signal which represents the neutron emission time from the target is the start time of time stamp. First sample point of the signal is the stop of time stamp. According to the start, stop time stamp and the time of signal over threshold, the total TOF can be obtained. One option is the time stamp add the threshold time is TOF. The other method is to use prompt gamma burst as the start of TOF, TOF is the time interval between the gamma burst signal and the other detector signal.

Back-n's flight length is limited to 80 m and the neutron pulse is widened by the thick target, so the time resolution at Back-n should not be high as neutron beam lines which is hundreds of meters. In view of the simulated neutron pulse width is nanosecond [11], sub-nanosecond time resolution is enough for TOF measurement.

## II. IMPLEMENT

The general-purpose readout electronics system is based on PXIe platform, which is composed of FDM (Field Digitizer Modules), TCM (Trigger and Clock Module), and SCM (Signal Conditioning Module). Fig2. Shows the general-purpose readout electronics system.

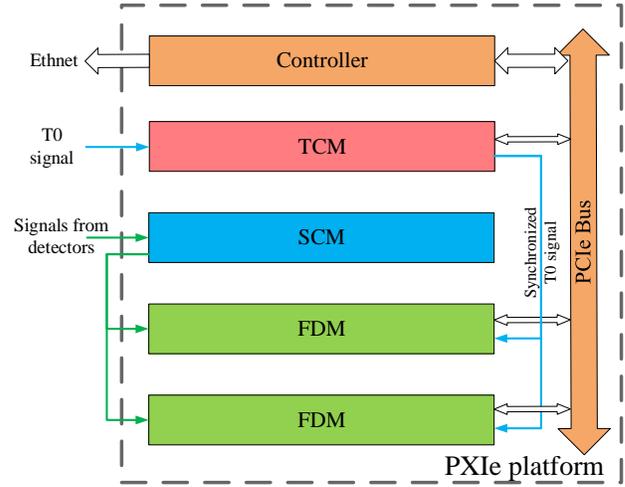

Fig. 2. The general-purpose readout electronics system

### A. Two methods of TOF measurement

T0 signal is generated by the front beam monitoring system and synchronous to the CSNS accelerator represents the neutron emission from the target. It is the start time of time stamp. The trigger and clock module (TCM) receives, synchronizes and distributes T0 signal to each FDM based on the PXIe backplane bus. Meantime, detector signal after being conditioned is fed into FDM for waveform digitizing, the first sample point of signal is the stop of time stamp, the time stamp add the time interval between the first sample point and threshold time is TOF.

Fig. 2 also shows the transmission paths of the T0 signal and the signal from detector. Fig. 3 shows the schematic of TOF in which T0 signal is the start of TOF.

The other method is to use prompt gamma burst as the start.

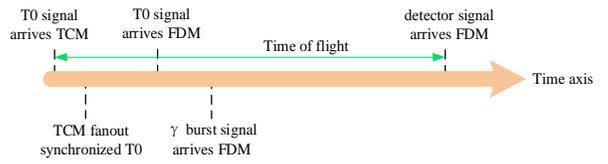

Fig. 3. TOF schematic when T0 is considered as the start signal

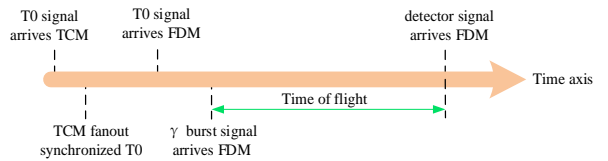

Fig. 4. TOF schematic when gamma burst is considered as the start signal

Every detector channel generates signal when prompt gamma burst arrives and there is a dedicate delay between the prompt gamma burst and neutron emission. By later correction, the prompt gamma burst can be used as the start of TOF. Fig. 4 shows the schematic.

### B. Receiving the asynchronous T0 signal

TCM is at the System Timing Slot of PXIe platform. When arriving TCM, T0 signal is an asynchronous signal to the general-purpose readout electronics system. The interval between T0 signal and FPGA Clock is recorded as t1. At the rising edge of TCM clock, TCM receives the T0 signal. After a few clock periods which is recorded as t2, TCM drives up to 17 synchronized T0 signal outputs to FDMs by differential star bus DSTARB in the platforms, FDMs are at the peripheral slots. Fig. 5 shows the transmission progress of T0 signal in TCM.

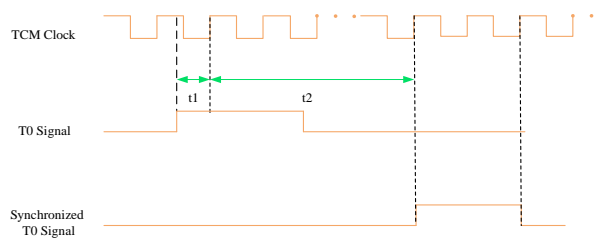

Fig. 5. The transmission progress of T0 signal in TCM

The interval between the T0 signal and TCM clock is measured by a FPGA-based TDC. In the readout electronics system, this interval will then be packed and uploaded to the chassis controller in DMA method via PCIe (Peripheral Component Interconnect Express) bus.

Though the clock of TCM and FDM is homologous, the phase is not deterministic every time the platform powers on. A FPGA-based TDC is used to measure the interval between T0 signal and FDM clock, this interval is recorded as t3, the architecture of this TDC is the same with the TDC employed in TCM.

Signals from detectors are conditioned by SCMs and then transmitted to FDMs, The ADC on FDM digitize the signal at 1G/s sampling rate. Count the time between the T0 signal synchronized with FDM and the first sampling data which is aligned with the clock, this interval is recorded as t4. The results and sampled data will then also be packed and uploaded to the chassis controller in DMA method via PCIe bus. The FDM results will be aligned with TCM results by T0 ID, time stamp which is the sum of t1, t2, t3 and t4 will also be aligned by T0 ID. Fig. 6 shows the diagram of t3 and t4.

The TDC employed on TCM and FDM are implied by carry chain [12]. Fig. 7 shows the architecture of TDC. In TCM and FDM, the input both are the asynchronous T0 signal. In TCM, the output of the first delay unit is named set signal, this signal generates at the first clock rising edge after T0 arriving and is used to get the UTC time. UTC time is a global reference time maintained by white rabbit system, TCM contains the interfaces with white rabbit system in the front board.

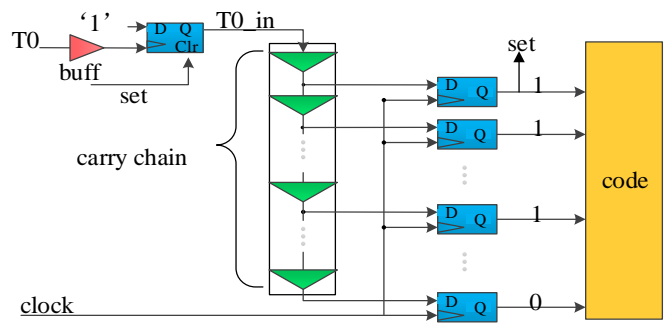

Fig. 7. The architecture of TDC

### C. Threshold time and the final TOF

Use CFD (Constant Fraction Timing) method to get the interval between the first data point and the threshold time after acquiring the data, this interval is recorded as t5. This method is used to evaluate the accuracy of TOF. In actually use, the timing method is optional because the signal are digitized and save in DAQ (Data Acquisition). Fig. 8 shows the schematic of

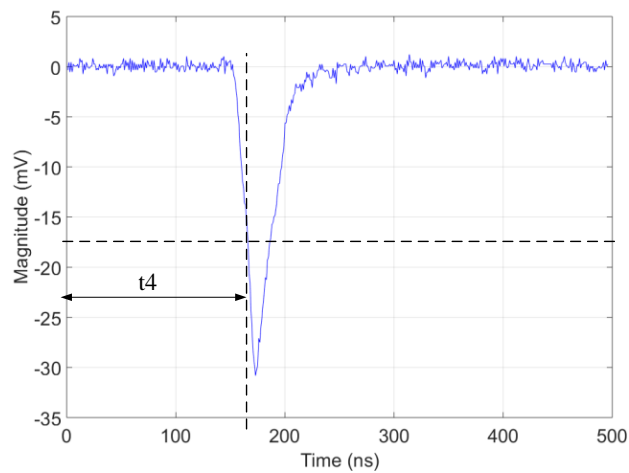

Fig. 8. The schematic of t5

t5. There are some dedicated intervals we don't elaborate, such as the transmission delay of the TCM, FDM and the back-plane buses, because they can be calibrated. Finally, the formula is showed as follow, in which d represents the sum of all the dedicated delays.

$$TOF = t1 + t2 + t3 + t4 + d$$

For the other method, the gamma burst signal is seen as the start of TOF' and the detector signal as the end. Use the same method to measure the interval of T0 signal and the detector signal, this interval is record TOF. So, TOF' is calculated as

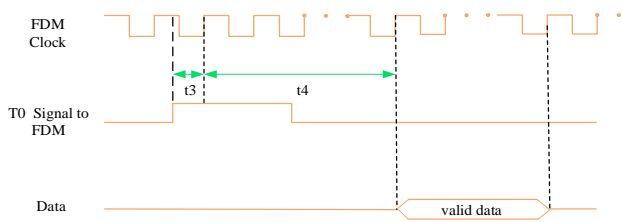

Fig. 6. The diagram of t3 and t4.

follow:
$$TOF' = TOF2 - TOF1$$

TOF1 is the TOF of the gamma burst and TOF2 is the TOF of the detector signal. The resolution reduce. Test results show the resolution of TOF measurement is enough for back-n.

This system is implemented on the general-purpose readout electronics, main modules of the electronics are showed in the Fig. 9.

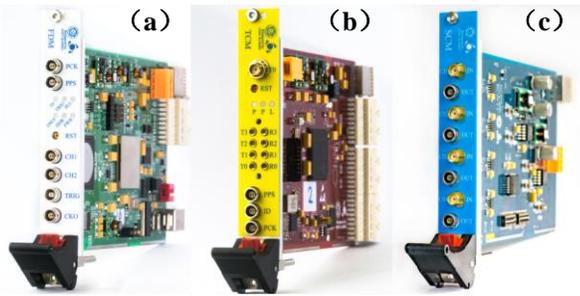

Fig. 9. The photograph of FDM (a), TCM (b), SCM (c)

## III. TEST

The Code density method is used to measure TDC's bin size. Signal source generates pulse signals asynchronous with TDC clock, the arriving time of pulse signals can be seen as uniform. For each delay unit, the number of pulse signals falling on represents the bin size.

The over-size bin is caused by cross-block carry chain. With the resolution requirement, bin-by-bin method is enough for revising [14]. The system clock is 125M and the max bin of TCM is 127. So, the average bin size is 63ps. The max bin of FDM is 174 and the average bin size is 46ps. The results are showed in Fig.10 and Fig.11.

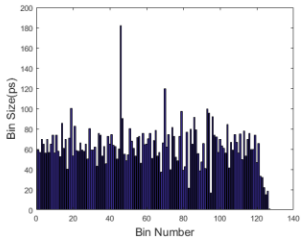 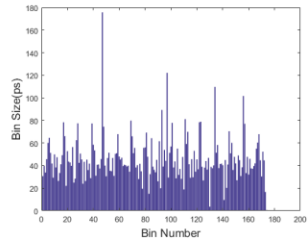

Figure 10. Bin size of TCM          Figure 11. Bin size of FDM

To evaluate the accuracy of the TOF measurement system, we simulate the T0 signal and the detector signal, use CFD method to find the capturing neutrons time offline, combine the results of TCM data and then get the TOF. Fig. 12 shows the test diagram of TOF. Signal Source generate two channel homologous signals. One is the T0 signal and the other is the detector signal.

As illustrated in Fig. 13, the RMS (Root Mean Square) of TOF is 280ps.

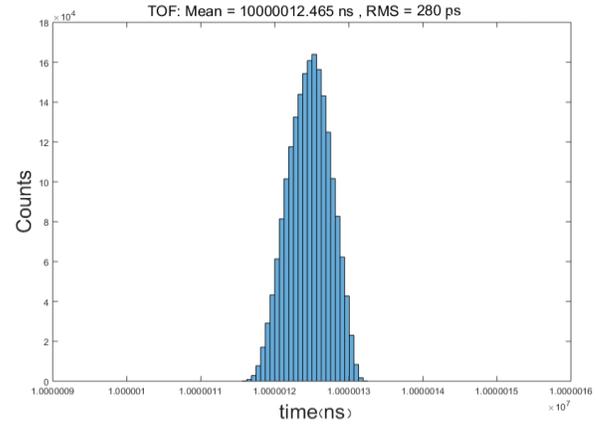

Figure 13. Accuracy of TOF

## IV. CONCLUSION

On the general-purpose readout electronics system of back-n, we have implemented a TOF measurement system to measure the neutron energy. All channels use the T0 signal as the reference signal, every signal has a time stamp, so they can be reestablished in one time axis. The time stamp is flexible for later data processing. The accuracy of the system is sub-nanosecond in large dynamic range of 10ms and applicable for Back-n. We will calibrate it in actually use.

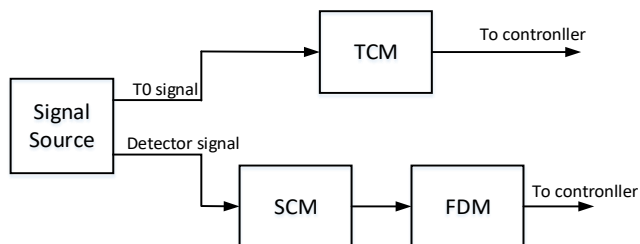

Fig. 12. Test diagram of TOF